\documentclass[prl,superscriptaddress,twocolumn,showpacs,amsmath,amssymb]{revtex4}
\usepackage[T1]{fontenc}
\usepackage{graphicx}% Include figure files
\usepackage{dcolumn}% Align table columns on decimal point
\usepackage{bm}% bold math
\usepackage{array}
\input epsf
\renewcommand{\vec}[1]{\mbox{\boldmath $#1$}}
\begin{document}
\preprint{}
\title{Impact of tensor force on $\beta$-decay of magic and semi-magic nuclei}
\author{F. Minato}
\affiliation{Nuclear Data Center, Japan Atomic Energy Agency,
Tokai 319-1195, Japan}
\author{C. L. Bai}
\affiliation{Department of physics science and technology, Sichuan University,
Chengdu 610065, China}
\date{\today}
\begin{abstract}
Effect of the tensor force on $\beta$-decay is studied in the framework of the
proton-neutron random-phase-approximation (RPA) with the Skyrme force.
The investigation is performed for even-even semi-magic and magic nuclei,
$^{34}$Si, $^{68,78}$Ni and $^{132}$Sn.
The tensor correlation induces strong impact on low-lying Gamow-Teller state.
In particular, it improves the $\beta$-decay half-lives.
$Q$ and $ft$ values are also investigated and compared with experimental data.
\end{abstract}
\maketitle
Nuclei far from the stability line is now one of highlighted topics in
the field of the nuclear physics.
Those nuclei are considered to have a characteristic surface structure
(e.g. neutron skin and halo) as well as a different shell structure
from stable nuclei.
Otsuka {\it et al.} indicated the importance of tensor force to explain
the shell evolution of neutron-rich nuclei \cite{Otsuka}. Later on,
studies of the tensor force toward neutron- and proton-rich nuclei
have been performed within the framework of the mean field
\cite{Lesinski2007,Brink2007,Bender2009,Suckling2010,Zhou2010}.
The recent analyses have shown that the tensor force improves the evolution
of empirical single-particle energies in $Z=50$ isotopes and $N=82$ isotones
\cite{Otsuka2006,Colo2007} and other nuclei with magic numbers \cite{Macij2008}.

In order to study the effects of the tensor force on the collective
excitation states in addition to the static properties, self-consistent
Hartree-Fock plus random phase approximation
(HF+RPA) schemes have been developed~
\cite{Bai1,Bai2,Cao2009,Donno2009,Donno2011,Co2009,Anguiano2011}.
Cao {\it et al.} have shown that the tensor force affects a magnetic
dipole excitation more sensitively than other multipole excitations
because of not only changes of the spin-orbit splitting but also its
RPA correlations. The Gamow-Teller (GT) and charge exchange
1$^+$ spin-quadrupole transitions in $^{90}$Zr and $^{208}$Pb are also studied
in Ref.~\cite{Bai1,Bai2}.

Apart from the stability line, the GT transition plays an important role
on $\beta$-decay of many unstable nuclei. Concerning it, the GT strength
and the corresponding $\beta$-decay half-lives ($T_{1/2}$) have been
investigated within a self-consistent proton-neutron quasi-particle
RPA (QRPA)~\cite{Engel1999,Niksic2005,Tomislov2007}.
In this approach, however, low-lying GT $1^+$ states of daughter nuclei
are produced at rather higher energy, and correspondingly the half-lives
become longer than the experimental data.
To remedy it, the phenomenological isospin $I=0$ proton-neutron ($pn$)
pairing is introduced in the QRPA, which plays an important role to
shift the low-lying GT peaks to a lower energy region and reproduce
isotopic and isotonic dependences of
$\beta$-decay half-lives \cite{Engel1999,Niksic2005,Tomislov2007}.
However, the effect of the $I=0$ $pn$ pairing has shown to play a minor
role in unstable closed- and sub-closed-shell nuclei, and the low-lying
GT states for those nuclei remain at a high energy.
Thus, the half-lives are longer than experimental data, {\it e.g.},
$^{132}$Sn becomes stable or extremely
long-life depending on adopted effective interactions
in contradiction to the experimental data ($T_{1/2}=39.7$ sec).

As a rule, systematical reproduction of known half-lives is required
for the reliable predictiction of unstable nuclei where one cannot
reach experimentally.
Such theoretical prediction is applied not only to the r-process
simulation which is the leading candidate synthesizing heavy elements
in the astrophysical sites but also the evaluation of decay-heat of
fission products yield. There might exist other candidates except $I=0$ $pn$
pairing that can reduce the energy of GT state and the half-life.
The tensor force, which has not been included in the previous QRPA
calculations, may alter them, considering the scattering of the GT
distribution found in the stable nuclei~\cite{Bai1,Bai2}.
In this letter, we clarify the role of the tensor force on
$\beta$-decay when the tensor force is included within the framework
of the proton-neutron RPA approach on the basis
of the Skyrme-HF method.

We use the triplet-even and triplet-odd zero range tensor terms
introduced originally by Skyrme\cite{Skyrme},
\begin{equation}
\begin{split}
v_t&(\vec{r}_1,\vec{r}_2)=
\frac{T}{2}
\left\{\left(
(\vec{\sigma}_1\cdot\vec{k}')(\vec{\sigma}_2\cdot\vec{k}')
-\frac{1}{3}(\vec{\sigma}_1\cdot\vec{\sigma}_2)\vec{k}'^2
\right)\delta(\vec{r})\right.\\
&+
\delta(\vec{r})\left.\left(
(\vec{\sigma}_1\cdot\vec{k})(\vec{\sigma}_2\cdot\vec{k})
-\frac{1}{3}(\vec{\sigma}_1\cdot\vec{\sigma}_2)\vec{k}^2
\right)\right\}\\
&+U
\left\{(\vec{\sigma}_1\cdot\vec{k}' ) \delta(\vec{r})
(\vec{\sigma}_2\cdot\vec{k})-\frac{1}{3}(\vec{\sigma}_1\cdot\vec{\sigma}_2)
\times (\vec{k}'\cdot \delta(\vec{r}) \vec{k})\right\},
\end{split}
\end{equation}
where $\vec{r}=\vec{r}_1-\vec{r}_2$, and the operator
$\vec{k}'=-(\vec{\nabla}_1-\vec{\nabla}_2)/2i$ acts on the left and
$\vec{k}=(\vec{\nabla}_1-\vec{\nabla}_2)/2i$ acts on the right.
The coupling constants $T$ and $U$ denote the strengths of the triplet-even and
triplet-odd tensor force interactions, respectively.
We restrict ourselves in spherical systems and solve the HF equation in
coordinate space.
We include the tensor force, the moment-independent, -dependent terms
and the spin-orbit terms in the p-h matrix element of the RPA equation
self-consistently.
In order to include continuum states, we discretize them with a box
boundary condition with the size of $16.0$ fm and a step of $dr=0.1$ fm.
We include the single-particle states up to $\epsilon_{\rm cut}=30$ MeV,
and truncate the RPA model space at the particle-hole energy of
$E_{\rm cut}^{\rm{p-h}}=70$ MeV. In this model space,
the Ikeda sum-rule $3(N-Z)$ \cite{Ikeda1964} is satisfied well for
all the nuclei that we studied.
In order to calculate $T_{1/2}$, $Q(\equiv Q_\beta-E_1^+)$ and $ft$ values,
we use same ansatz as Sec.II of Ref. \cite{Engel1999} with the ratio
of vector and axial-vector constants $g_A/g_V=1.26$ in the weak interaction,
where $E_{1^+}$ is the energy of GT $1^+$ state of daughter nuclei.

To evaluate $\beta$-decay in the present framework, an appropriate
effective interaction must be used. As discussed in Ref. \cite{Engel1999},
the distribution of GT strength depends primarily on the time-odd part
of the Skyrme energy density, while usual Skyrme parameter
sets are fitted by the ground state properties of even-even nuclei
discribed by the time-even one.
As a result, most of the Skyrme parameter sets do not assure of
applying it to the calculation of $\beta$-decay. 
Engel {\it et al.} assessed the strength of GT distribution with various
Skyrme interactions.
They argued that the s-wave time-odd Landau-Migdal parameter
$\mathrm{g}_0'$ is one of the scales to select an appropriate
interaction giving a reliable low-lying GT distribution \cite{GS1981}.
They found that it must not be a small value around zero and finally
adopted SkO' interaction ($\mathrm{g}_0'=0.79$) \cite{Reinhard1999} to
apply the self-consistent QRPA to the calculation of $\beta$-decay
half-lives \cite{Engel1999}.
Bender {\it et al.} performed a more detail examination of several major
Skyrme parameter sets, especially SkO' as well as Engel {\it et al.}
to study the GT resonance for the future application to
$\beta$-decay calculation \cite{Bender2002}.
However, the perturbation of tensor force on SkO' has not been studied well,
and no tensor force is proposed on SkO'.
We noticed SkO interaction ($\mathrm{g}_0'=0.98$),
which is determined in the same manner as SkO',
but neglects the $\vec{J}^2$ terms of the exchange part of
the central force ($A_t^J$ terms in Ref. \cite{Lesinski2007})
in the mean-field in the evaluation.
SkO provides GT distribution at a comparable level to SkO'
systematically \cite{Bender2002}. 
On the top of SKO, tensor force
is added perturbatively with T=553.7 and U= -325.7 fm$^5$MeV, 
and this effective interaction is called SKO$_{T^\prime}$.
The spin-orbit strength of SkO$_{T'}$ is reduced by $15\%$
when the tensor force is added,
however this reduction does not affect our conclusion.

In addition the above studies, Fracasso {\it et al.} 
discussed the effect of the $\vec{J}^2$ terms on the GT resonance
and isovector spin-dipole \cite{Fracasso2007}.
They argued that use of $\mathrm{g}_0'$ around $0.45\sim0.5$ is reasonable,
not around empirical value evaluated from the Woods-Saxon potential
($\mathrm{g}_0'\sim \mathrm{1.8}$),
in order to reproduce a proper proportion of the strength exhausted
by giant GT resonance. Take this argument into account, we noticed that
Skx, of which $\mathrm{g}_0'=0.51$~\cite{BA.Brown1998}.
With the same manner of fitting Skx, 
a new interaction Skxta is fitted with tensor force 
($T=384.0$, $U=144.0$ fm$^5$ MeV) \cite{BA.Brown2006}.

Apart from the two interaction, we considered other several major interactions,
SIII ($\mathrm{g}_0'=0.95$), SGII ($0.93$), SLy5 ($-0.15$) and T43 ($0.14$).
We referred to SIII+Tensor \cite{Brink2007}, SGII+Te1 \cite{Bai2011} and
SLy5+Tensor \cite{Colo2007} for the strengths of the tensor
force of SIII, SGII and SLy5, respectively.
In carrying out RPA calculation with SkO, SIII and SGII,
the $A_t^J$ terms are taken into account in the residual interactions.
Although this prescription breaks self-consistency of our framework,
it is reasonable to sustain relevant position of GT resonance
\cite{Fracasso2007}.
And, when we include the tensor force on the top of Skx,
we keep parameters of central force remain that of Skx, not Skxta.
This is because the central part of Skxta is substantially different 
from that of Skx and use of Skxta makes it complicated to 
illustrate the effect of tensor force straightforwardly.
\begin{figure}
\begin{center}
\includegraphics[width=0.95\linewidth]{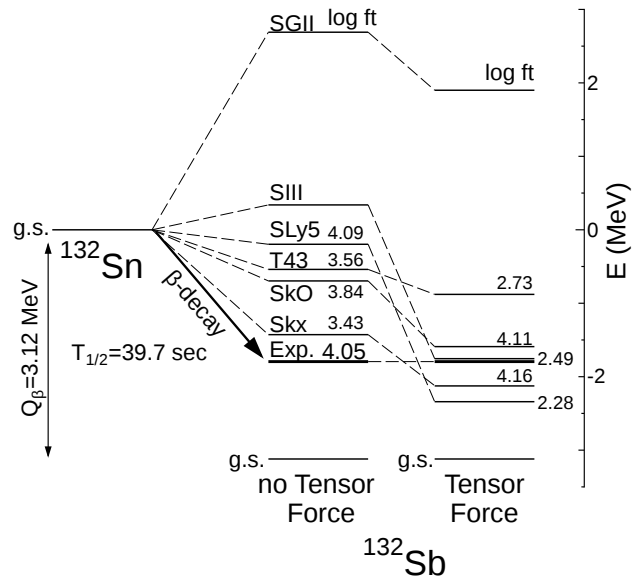}
\end{center}
\caption{Decay scheme of $^{132}$Sn with and without tensor force.
Calculated and experimental $1^+$ state of $^{132}$Sb
is shown by the thin and thick line, respectively.
The vertical axis on the right indicates energy with
respect to the ground state of $^{132}$Sn.}
\label{lgt}
\end{figure}

We first study the $Q_\beta$ and log $ft$ values of $^{132}$Sn, 
that is a typical spherical closed-shell nuclei 
in which the effect of pairing interaction is negligible.
In the previous theoretical studies it is stable or extremely long-life.
Figure \ref{lgt} shows the decay scheme of $^{132}$Sn.
Experimental $1^+$ state of daughter nucleus $^{132}$Sb decaying from
the ground state of $^{132}$Sn is shown by the thick line
and calculated $1^+$ state by the thin line.
Looking at the case without the tensor force,
SIII and SGII produce the $1^+$ states higher than the ground state
of $^{132}$Sn, consequently $\beta$-decay is not allowed.
The $1^+$ states of SLy5, T43, Skx and SkO are below the g.s. of $^{132}$Sn,
but they are still higher than the experimental data.
Including the tensor force, all the $1^+$ states are reduced systematically.
This is because the tensor force works attractively
in the isospin $T=0$ spin-parity $J^\pi=1^+$ channel at low energy region
in common with well known effect of tensor force on binding state of deuteron.
We should remark that the tensor force in this channel
works repulsively at high energy \cite{Bai1,Bai2}.
SGII and T43 with the tensor force are still so higher than
experimental data that they are not suitable for the present study,
while SIII, SLy5, SkO and Skx become close to experimental $1^+$ state.
However, SIII and SLy5 yield rather small $\log ft$ values
because they attract too strong GT strength at the lower energy.
As a consequence, 
we use SKO and Skx to discuss $\beta$-decay of magic and semi-magic nuclei
in the following.

\begin{figure}
\begin{center}
\includegraphics[width=0.40\linewidth]{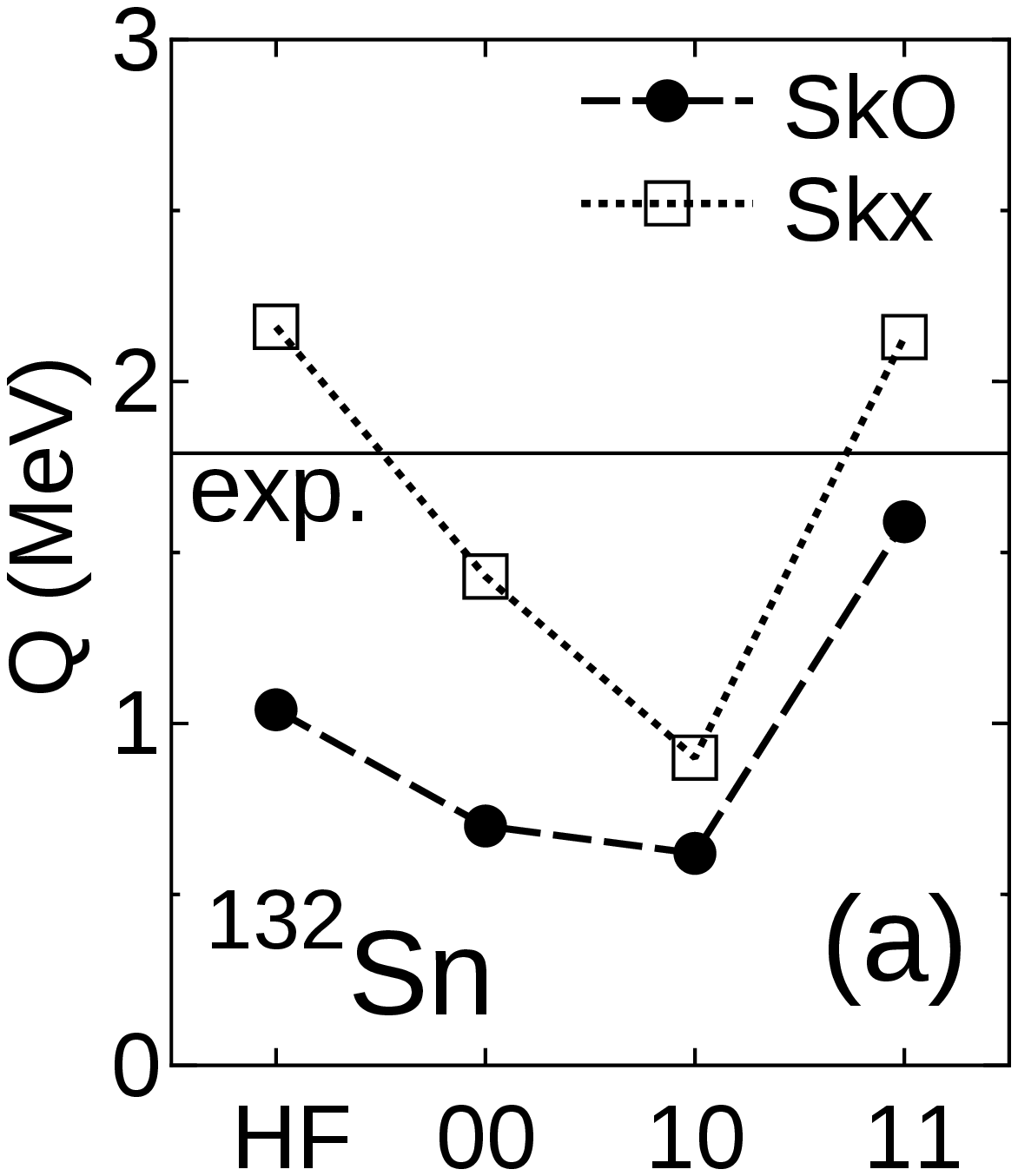}
\includegraphics[width=0.40\linewidth]{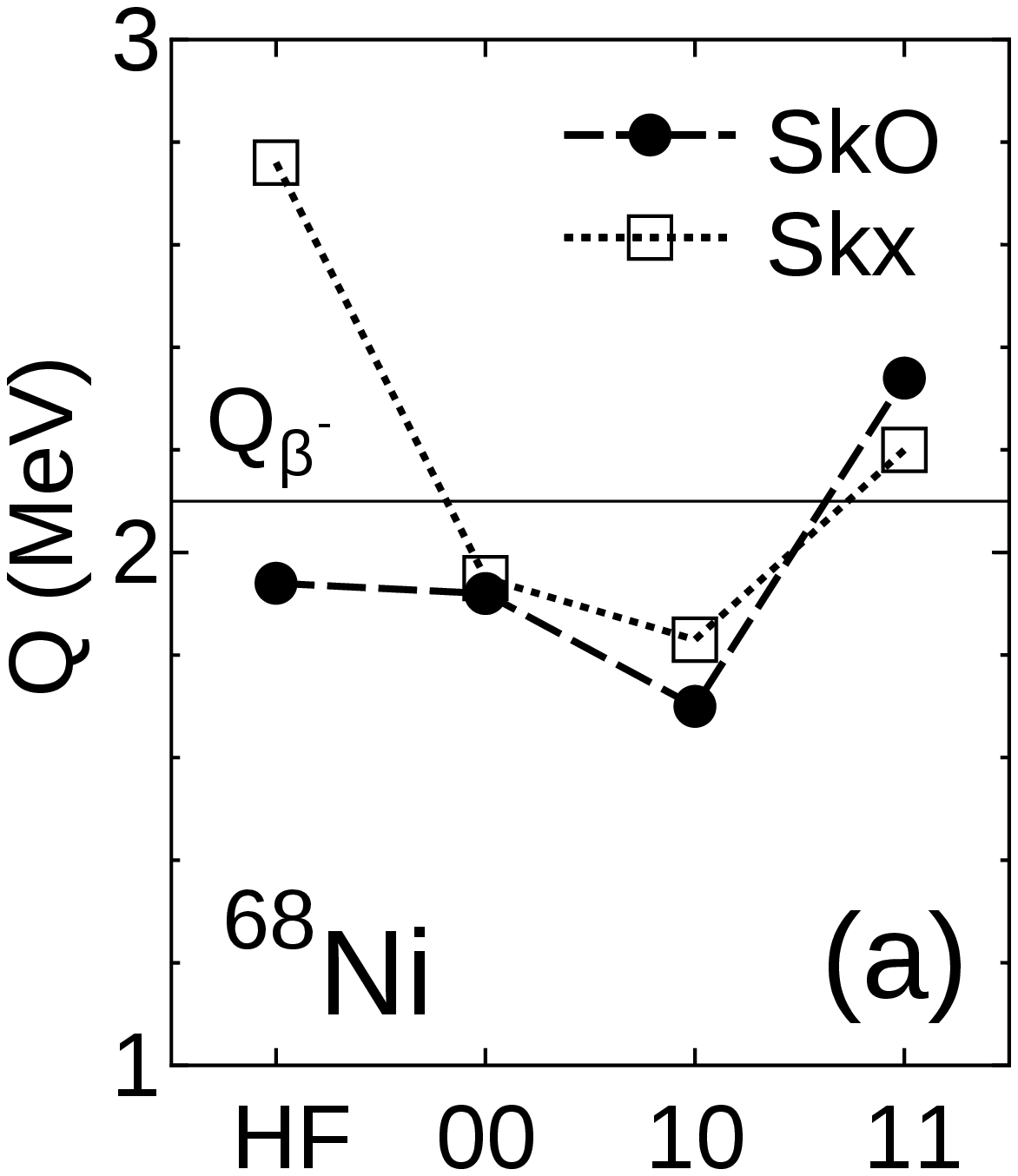}
\includegraphics[width=0.40\linewidth]{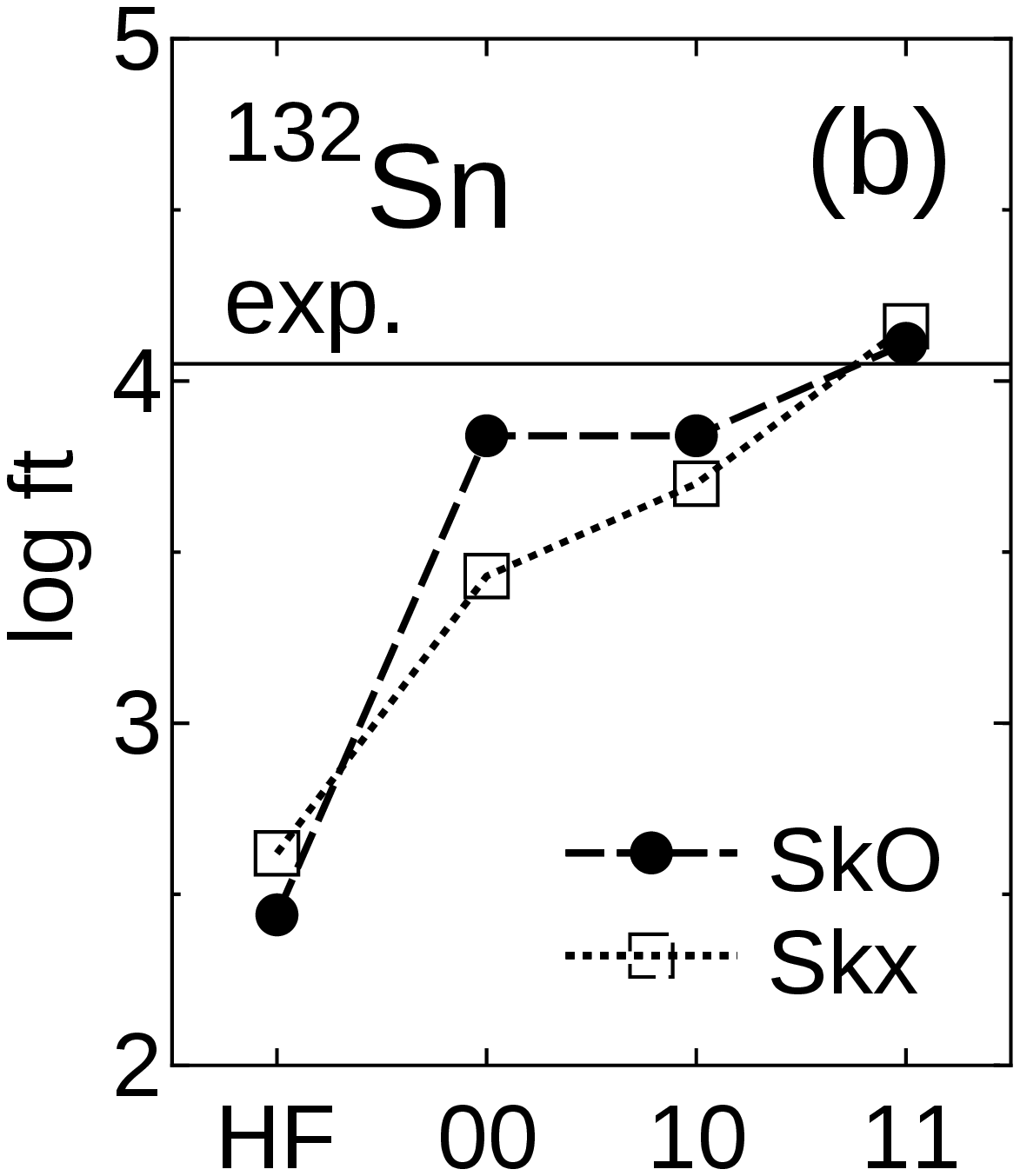}
\includegraphics[width=0.40\linewidth]{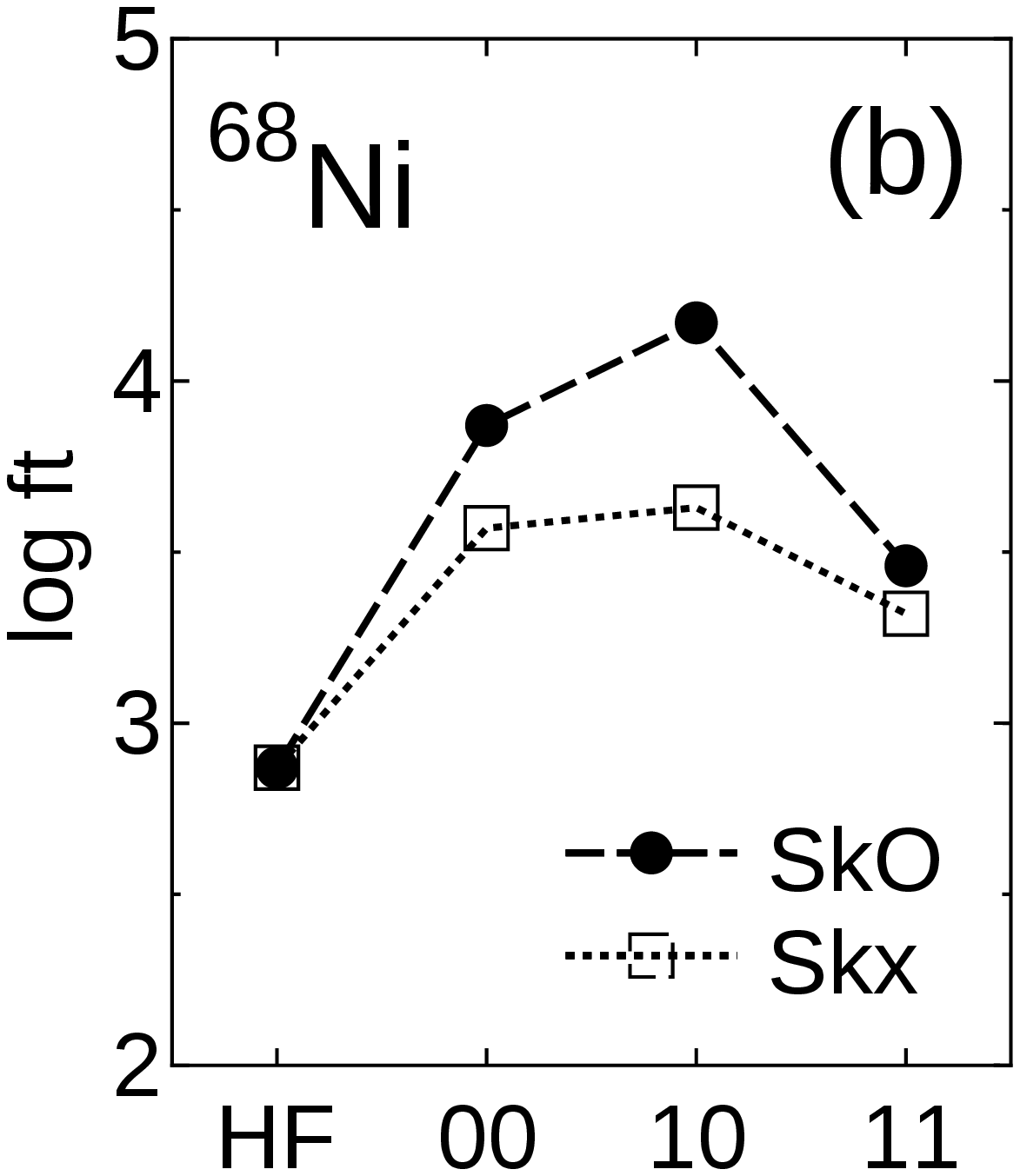}
\end{center}
\caption{$Q$ (panel a) and $\log ft$ values (panel b) of
the $\beta$-decay of $^{132}$Sn (left) and $^{68}$Ni (right).
Experimental data of $Q\equiv E_{1^+}-E_{g.s.}$ and $\log ft$
values of $^{132}$Sn and experimental $Q_{\beta^-}$ of $^{68}$Ni
\cite{Nudat} are shown with the holizontal line.
The index HF means the result without the residual interaction.
The indieces 00, 10 and 11 means the result of the HF+RPA without
the tensor force, only with the tensor force in the HF
and with the tensor force fully included in the HF and the RPA,
respectively.}
\label{qft}
\end{figure}

The contribution of the tensor force to the HF+RPA is mainly devided
to two parts.
One is the spin-orbit potential in the HF and another is the residual
interaction in the RPA which has not been included in the HF potential.
In order to analyze their contributions to the $\beta$-decay,
we consider three cases in the same way as Ref. \cite{Bai1}.
The first one is obtained without the tensor force both in the HF
and the RPA (labeled by 00).
The second one is obtained only with the tensor force in the HF,
{\it i.e.} in the spin-orbit splitting (labeled by 10).
In the last one, the tensor force is included fully in the HF and
the RPA (labeled by 11).
We discuss double magic and semi-magic nuclei, $^{132}$Sn,
$^{68,78}$Ni and $^{34}$Si in the present study.

Figure \ref{qft} shows the $Q\equiv E_{1^+}-E_{g.s.}$ (upper panel) and
$\log ft$ (lower panel) values of $^{132}$Sn and semi-magic nucleus
$^{68}$Ni where the pairing also plays a minor role as well as $^{132}$Sn.
We pay attention to the predominant transition which leads to the
shortest partial half-life. The holizontal line for $^{132}$Sn is
the experimental data. Unfortunately, any decay scheme of $^{68}$Ni
is not identified experimentally at present. Instead, experimental
$Q_{\beta^-}$ value \cite{Nudat} is shown by the holizontal line in the panel.
The $Q$ values for both forces are
shifted downward when only the central force is taken into account in
the residual interaction ({\it i.e.} 00) as compared to the case HF.
The same result is seen in $^{68}$Ni.
This is a general consequence because the central part
in the $1^+$ GT channel works repulsively due to the positive $g_0'$.
Introducing the tensor force in the spin-orbit splitting ({\it i.e.} 10),
the $Q$ values are further reduced and apart from experimental $Q$.
As the tensor force is fully included in the HF+RPA ({\it i.e.} 11),
the results of both SkO and Skx increase significantly and
become closer to the experimental value.
Similarly, the $Q$ value of $^{68}$Ni are shifted downward from the
case 00 to 10 and only the tensor terms in the residual interaction
shift them upward in the case 11.
The increase of $Q$ value for these two nuclei is therefore
mainly caused by tensor terms of the residual interaction,
which works attractively opposite to the central terms,
rather than the change of spin-orbit splitting.

Besides the $Q$ values, an improvement of $\log ft$ values is also obtained.
For the case HF, the calculated $\log ft$ values badly
underestimates the experimental one and out of typical value of
GT transition ($\log ft=3\sim6$) for both forces.
In the case 00, the results are improved both for $^{132}$Sn and $^{68}$Ni
because $\log ft$ value is inproportional to GT strength, which is
reduced by a coherent excitation ({\it i.e.} increase of collectivity)
evoked by the residual two-body interaction.
By inclusion of the tensor force in the HF ({\it i.e.} 10) and the
HF+RPA ({\it i.e.} 11), further improvements are obtained in the
result of $^{132}$Sn.
Although there are no experimental data, $\log ft$ values of
$^{68}$Ni settle around $3.5$, which agrees with the typical value.

\begin{figure}
\begin{center}
\includegraphics[width=0.40\linewidth]{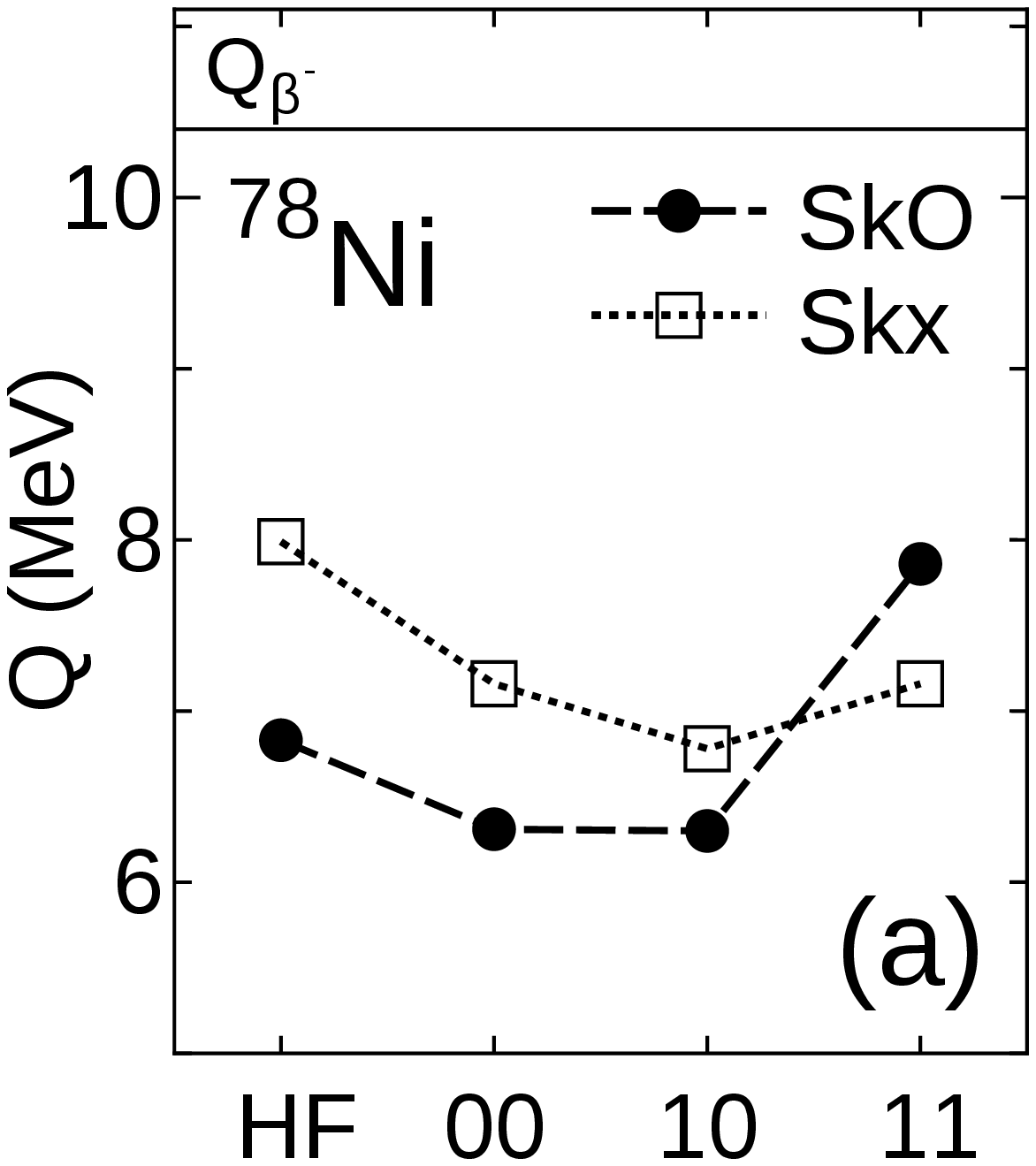}
\includegraphics[width=0.40\linewidth]{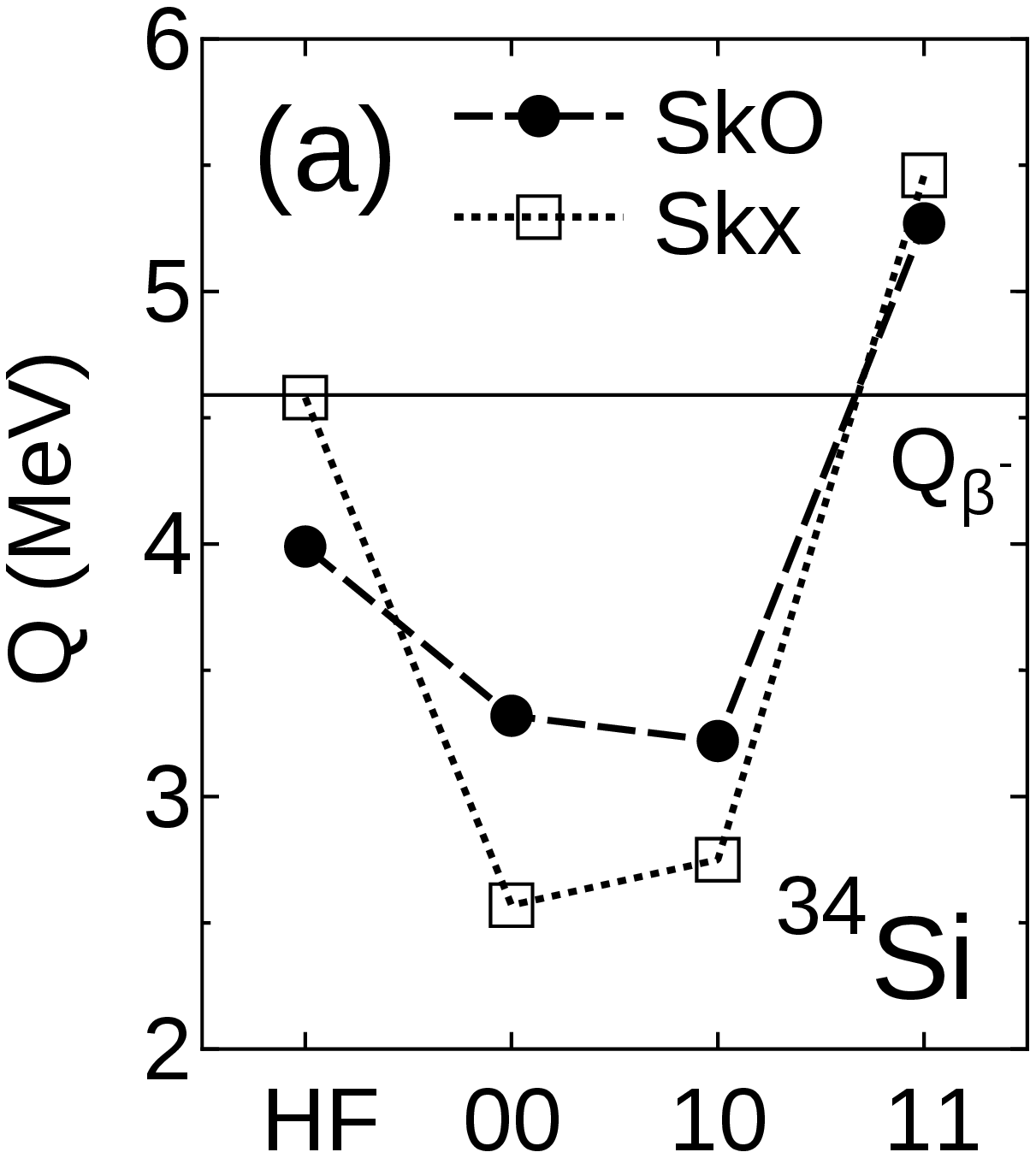}\\
\includegraphics[width=0.40\linewidth]{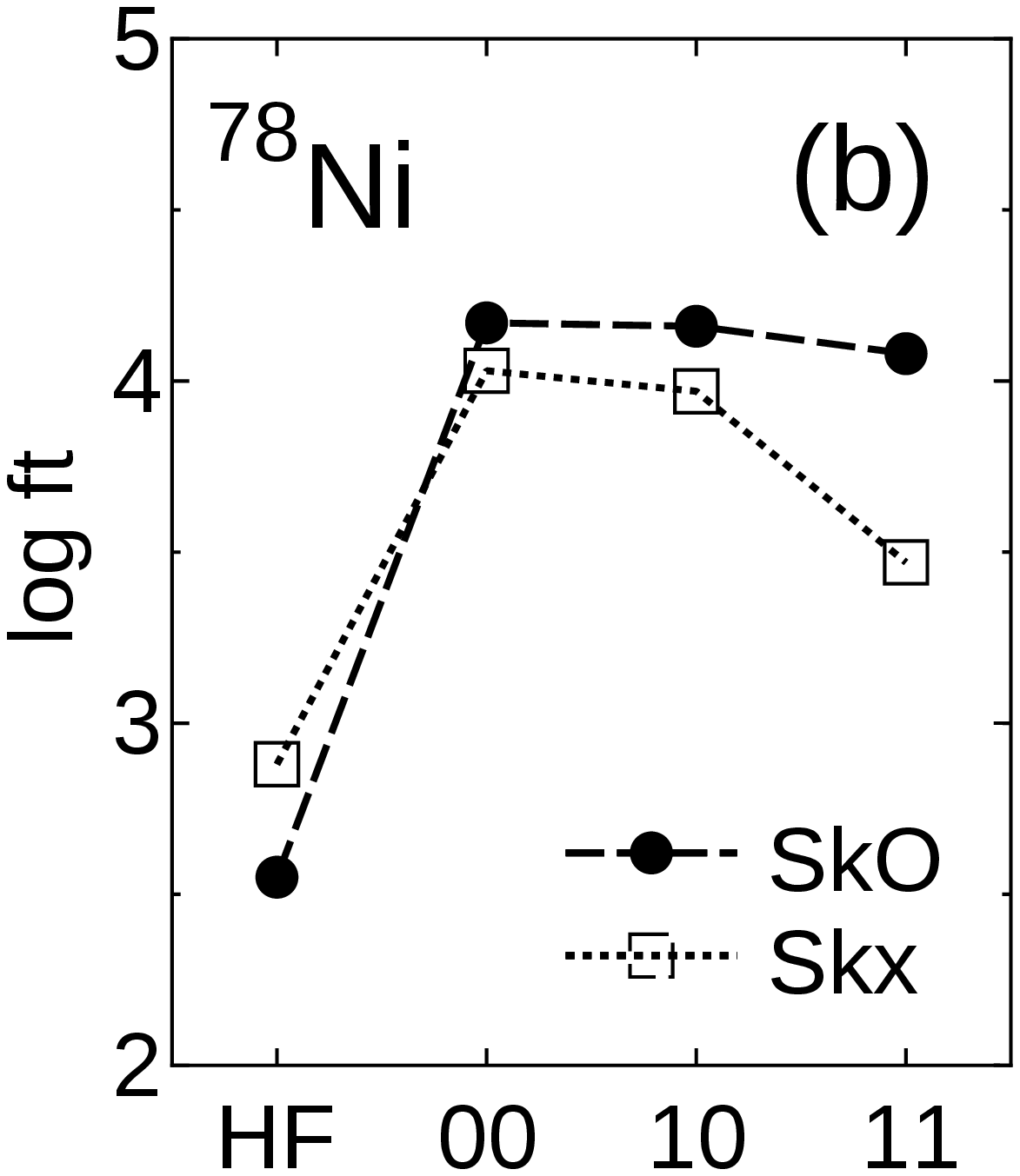}
\includegraphics[width=0.40\linewidth]{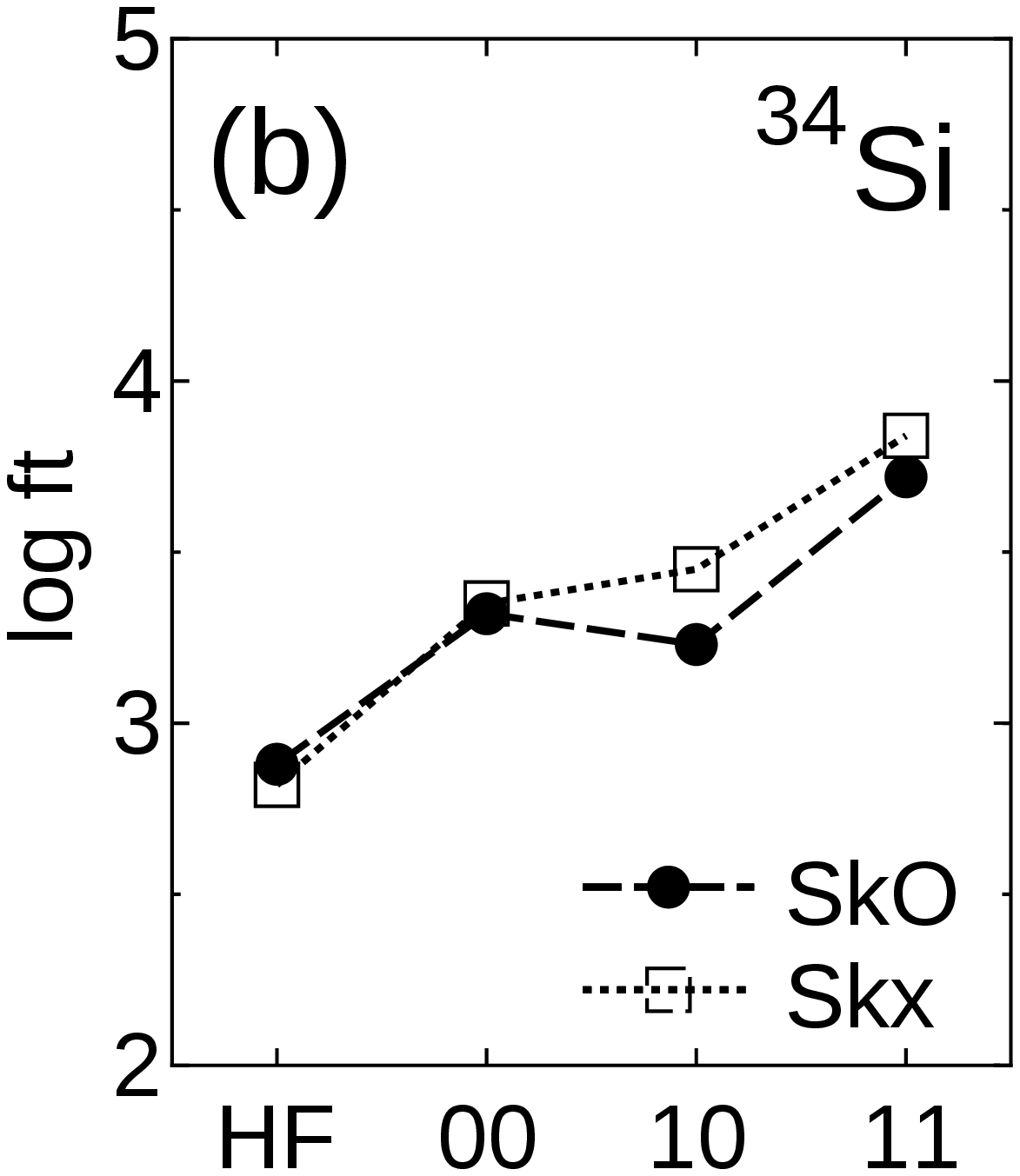}\\
\end{center}
\caption{Same as Fig.\ref{qft}, but for $^{78}$Ni (left) and $^{34}$Si (right).
experimental $Q_{\beta^-}$ of $^{78}$Ni and $^{34}$Si are
shown with the holizontal line.}
\label{qft2}
\end{figure}

Figure \ref{qft2} shows the $Q$ and $\log ft$ values of double-magic
nucleus $^{78}$Ni and semi-magic nucleus $^{34}$Si. We again pay
attention to the predominant transition which leads to the shortest
partial half-life.
Since the $Q$ values are not measured, experimental $Q_{\beta^-}$
values \cite{Nudat} are shown by the holizontal line, instead.
The similar dependence of central and tensor forces to $^{132}$Sn
and $^{68}$Ni is obtained. The $Q$ value shifts downward from the case HF to 00. 
Although the tensor terms in the spin-orbit splitting also assists 
the slight increase of $Q$ value in case of $^{78}$Ni(SkO) and $^{34}$Si(Skx),
the tensor terms in the residual interaction again shift $Q$ values
the most meaningfully.
The $\log ft$ values of $^{78}$Ni and $^{34}$Si settle around $3.5\sim4$
as a result of including the tensor force in the case 11 
from the edge of typical value ($\log ft \sim 3$) in the case HF.

\begin{figure}
\begin{center}
\includegraphics[width=0.49\linewidth]{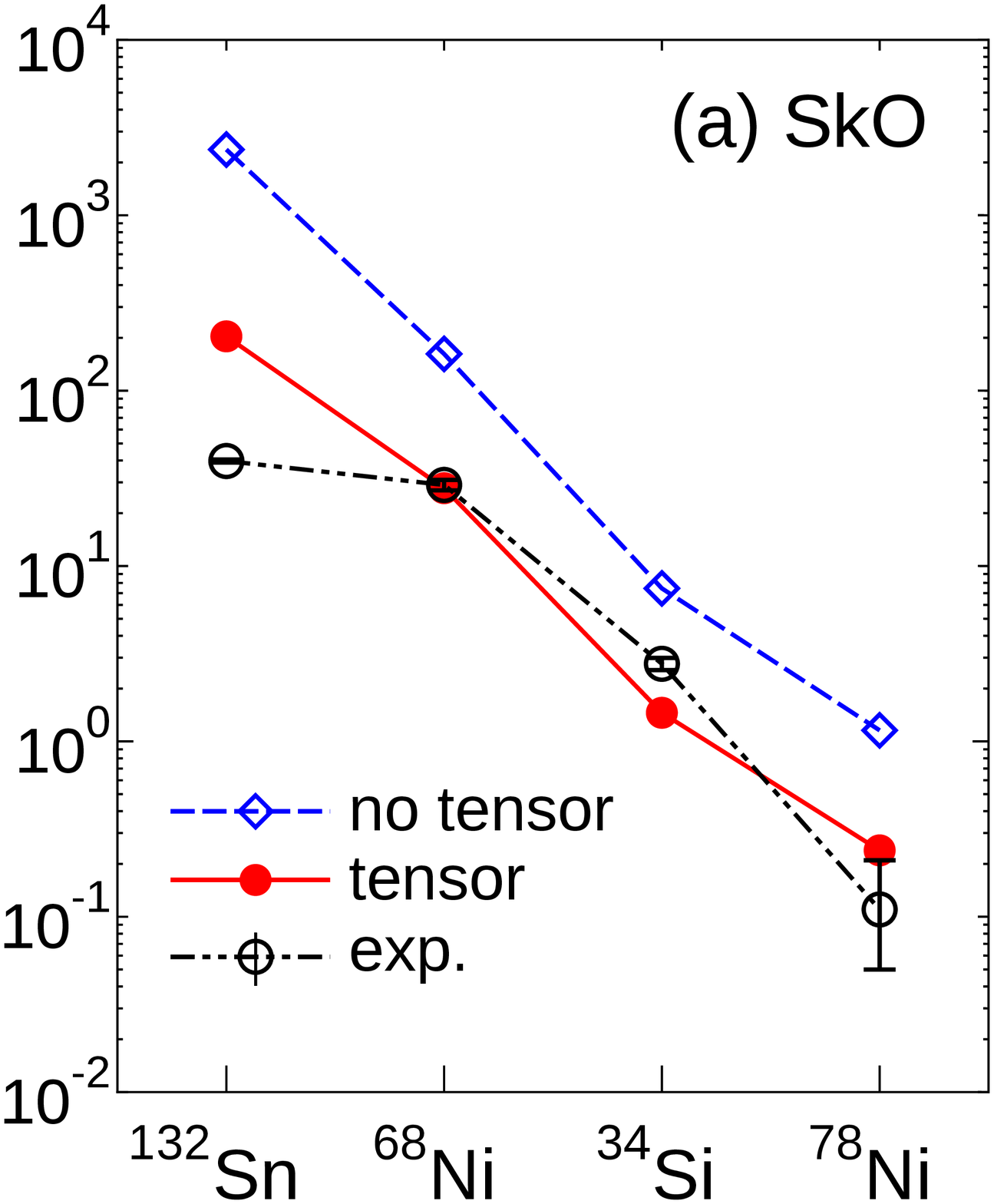}
\includegraphics[width=0.49\linewidth]{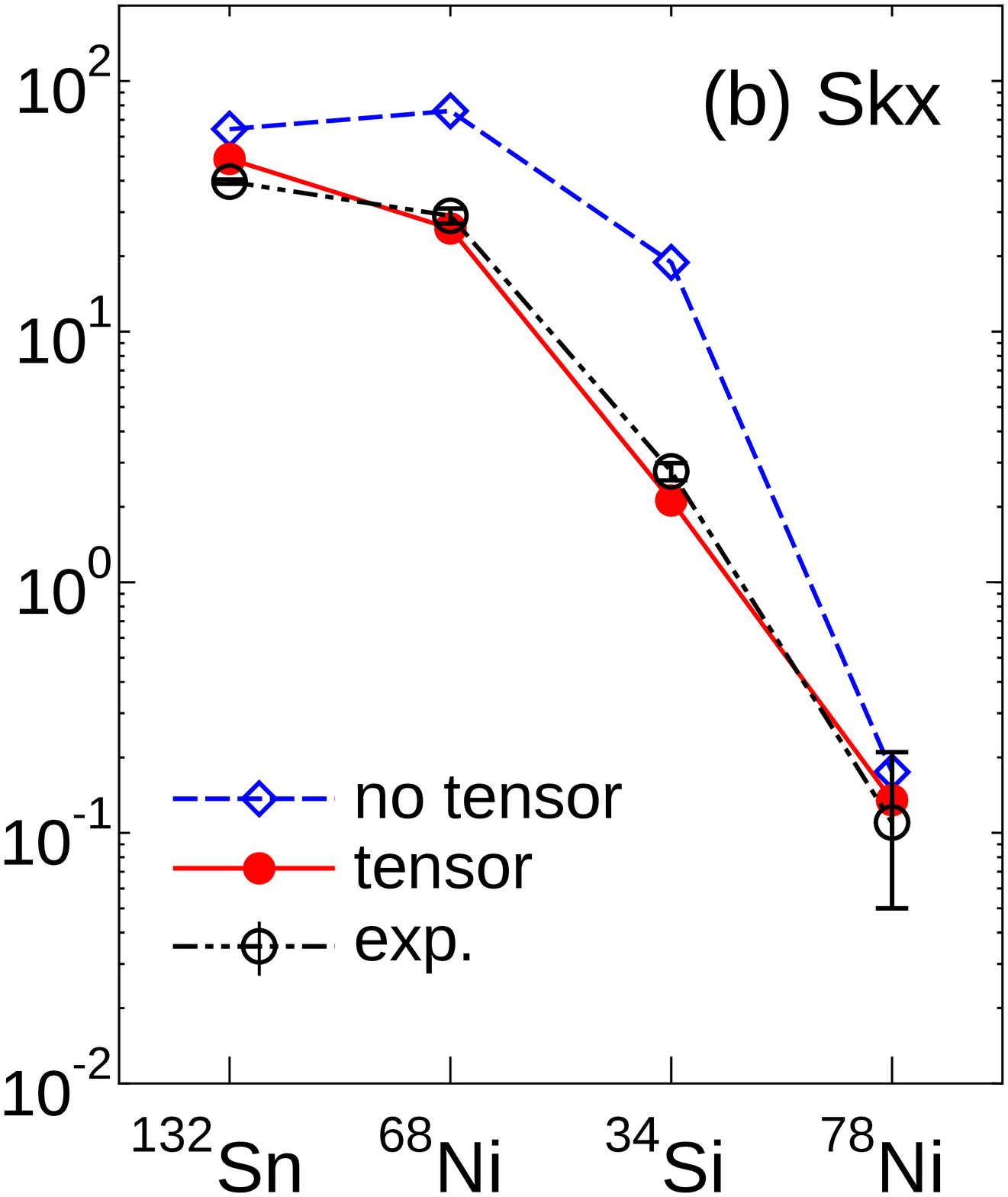}
\end{center}
\caption{(Color online) $\beta$-decay half-lives with and without
the tensor force of $^{132}$Sn, $^{68}$Ni, $^{34}$Si and $^{78}$Ni
(in order of half-life) calculated by SkO (a) and Skx (b).
Experimenta data are taken from Ref. \cite{Nudat}.
The vertical axis is given in the unit of second.}
\label{hlf}
\end{figure}

Figure \ref{hlf} shows the $\beta$-decay half-lives of $^{132}$Sn,
$^{68}$Ni, $^{34}$Si and $^{78}$Ni (in order of half-life).
The diamond and filled circles indicate the results without
the tensor force and with tensor force fully in the HF+RPA, respectively.
The experimental data is indicated by the open circle.
The result without the tensor force overestimates the
experimental data systematically.
On the other hand, the calculation with the tensor force
shows better agreements with the experimental data for both forces.
As seen, this is mainly because the GT strengths are attracted to 
the lower energy by the tensor force and consequently
the $Q$ value, which was small in
case without the tensor force, increases meaningfully.

We discussed the role of the tensor force on the $\beta$-decay 
by including tensor force in HF+RPA methods.
We found that tensor force makes dramatic improvement in predicting
the $\beta$-decay half-lives of the even-even
semi-magic and magic nuclei $^{132}$Sn, $^{68}$Ni, $^{34}$Si, and $^{78}$Ni
where the effect of Iso-scalar paring is negligible. 
The tensor force gave a significant contribution to the low-lying GT
distribution, in particular, the effect of the tensor terms in
the residual interaction of RPA was more important than the change
of the spin-orbit splitting in the HF.
Of course, there are still some points to be discussed for the
precise prediction of $\beta$-decay with the present approach,
{\it e.g.}, time-odd part of the effective interaction,
effect of 2p-2h state, etc. The study of tensor effect on the
GT distribution just started toward its devlopment.
Nevertheless, the effect of the tensor force on the low-lying
GT peaks is obvious so that it is requested to include it for
the accurate and reliable prediction of the $\beta$-decay half-life
in the self-consistent RPA and QRPA approaches.
We intend to develop our formalism to open-shell and/or deformed nucleus
for further investigation of the tensor force effect on the beta-decay
for future.

We thank H. Sagawa in Univ. of Aizu for useful discussions.

\end{document}